\documentclass[11pt,twoside]{article}

\usepackage{asp2006}
\usepackage{epsf}
\usepackage{lscape}
\usepackage{natbib}
\usepackage{graphicx}

\begin{document}
\title{Binaries among Extreme Horizontal Branch Stars in Globular Clusters}
\author{C. Moni Bidin,$^1$ M. Catelan,$^2$ S. Villanova,$^3$
  G. Piotto,$^3$ M. Altmann,$^4$ Y. Momany,$^5$ and S. Moehler$^6$}
\affil{$^1$Departamento de Astronom\'{i}a, Universidad de Chile,
Casilla 36-D, Santiago, Chile}
\affil{$^2$Pontificia Universidad Cat\'{o}lica de Chile,
Departamento de Astronom\'{i}a y Astrof\'{i}sica, Av. Vicu\~{n}a Mackenna
4860, 782-0436 Macul, Santiago, Chile}
\affil{$^3$Dipartimento di Astronomia, Universit\`{a} di Padova,
Vicolo dell'osservatorio 3, 35122 Padova, Italy}
\affil{$^4$University of Heidelberg, Centre for Astronomy,
  M\"{o}nchhofstr. 12-14, 
D-69120 Heidelberg, Germany} 
\affil{$^5$INAF-Osservatorio Astronomico di Padova, 
Vicolo dell'osservatorio 2, 35122 Padova, Italy}
\affil{$^6$European Southern Observatory, Karl-Schwarzschild-Str. 2,
85748 Garching, Germany}

\begin{abstract}
Following a review of our present knowledge about blue subdwarf
stars in globular clusters, we present an overview of the
results of searches for close binaries among these stars, including
results previously published in the literature and reporting recent and
preliminary results of new data.
Previous investigations revealed a lack of close systems in NGC\,6752,
which we confirm with new, more extensive observations. Our estimate of
the close binary fraction in this cluster is only 4\%. From a review of
the relevant literature, there are indications that a low close binary
fraction among extreme horizontal branch (EHB)
stars is a common feature in globular star clusters.
On the other hand, the field EHB population shows evidence of a remarkably
high binary fraction. Such a difference among globular cluster and field
populations, although not yet explained in detail on the basis of theoretical
models, must necessarily be related to different formation histories for EHB
stars in the field and in clusters. In this framework, preliminary
results indicate that close systems could be relatively common in the
peculiar globular cluster NGC\,2808, although the sample of studied stars
is still small.
This would imply that not all clusters share the same behavior, as far as
EHB star formation is concerned. We briefly explore possible explanations
for these results.
\end{abstract}


\section{sdB Stars in Globular Clusters}

The horizontal branch (HB) is a well-known feature in the
color-magnitude diagram (CMD) of Galactic globular clusters (GCs),
first discovered by \citet{Bruggencate27} and later identified as a
population of stars which, after the evolution along the red giant
branch (RGB), eventually ignite He-burning in the core
\citep{Hoyle55}.  \citet{Greenstein71} and \citet{Caloi72} identified
the hotter ($T_\mathrm{eff}\geq \,20,\!000$~K) HB extension with the
already known field B-type subdwarf (sdB) stars, which as a
consequence are also referred to as extreme horizontal branch (EHB)
stars. 

Because of their intrinsic faintness at optical wavelengths, sdBs in
GCs were hardly observed at first \citep{Cox61}, and only later did
it become evident that they were present in many GCs, even at high
metallicities \citep[e.g.][]{Rich97}. Thus they became the most prominent
evidence of the well-known {\it second parameter problem}, i.e. the
large degree of variation in HB morphology partly explained by
metallicity \citep[the ``first parameter'';][]{Sandage60} but not
completely accounted for by it \citep{Sandage67,VanDenBergh67}.
\citet{Rood73} pointed out
that the spread in temperature along the HB implies a corresponding
spread in envelope mass, bluer stars being less massive, and EHB stars
must have lost almost their entire envelope during their
evolution. The retained envelope is so thin (about $0.02\,M_\odot$)
that it is insufficient to sustain later evolution along the
asymptotic giant branch (AGB), and after He exhaustion in the core EHB
stars are predicted to evolve directly to the white dwarf cooling
sequence \citep[see][for a recent review of low-mass stellar
evolution]{Catelan07d}. Investigations have accordingly mainly been
focused on finding the mechanism(s) responsible for the heavy RGB mass
loss required for EHB formation.

\subsection{Models for EHB Star Formation}
\label{introformation}

Many canonical and non-canonical models have been proposed over the past several
decades, either considering EHB stars as distinct populations or just as the hotter components
of a continuous HB. Some very recent results allow us to focus our attention on some of these
scenarios \citep[see also][for a discussion in the context of the ultraviolet upturn phenomenon
in galaxies]{Catelan07a}.

The recent discovery of a planet orbiting a field sdB star \citep{Silvotti07} adds
new emphasis on the hypothesis that a sub-stellar body interacting with the expanding
envelope of an RGB star can cause heavy mass loss, thus leading to the formation of
an sdB star \citep{Soker98}. Searches for planets around main sequence (MS) stars
in GCs have been fruitless thus far \citep{Gilliland00,Weldrake07}; indeed,
\citet{Soker07} argue that similar such searches are hopeless, because of the long
predicted orbital periods.

A super-solar He abundance in the stellar envelope can enhance the
mass loss along the RGB by increasing the RGB tip luminosity, thus
leading to the formation of an sdB star. Non-canonical phenomena were
invoked including He mixing driven by internal rotation
\citep{Sweigart79,Sweigart97} and dredge-up induced by H-shell
instabilities \citep{VonRudloff88}, but the recent discovery of
He-enriched sub-populations in $\omega$~Cen
= NGC\,5139 \citep{Bedin04,Piotto05}
and probably NGC\,2808 \citep[as suggested by][]{Piotto07,DAntona05}
suggests that the He enhancement could be
primordial. Such He enhancement would also affect the HB
morphology \citep[e.g.,][]{Lee05,DAntona05}.  Unfortunately, a He enhancement
would have
little consequence on the stellar surface parameters in this
high-temperature regime, independently of its origin, and observed He
abundances are altered by atmospheric phenomena like gravitational
settling.  Therefore, spectroscopic measurements cannot help decide
between the EHB progeny of He-enriched or He-normal stars
\citep[e.g.,][]{Moehler00,Moni07}. Note that there do exist even hotter stars
along the extension of the HB, the so-called ``blue hook stars''
\citep[e.g.,][]{DCruz00}, which will be discussed below. To the best of
our knowledge, the crucial phenomenon suspected to be responsible for
their formation has yet to be properly incorporated into the CMD
simulations that are usually carried out in the framework of the
increased He scenario.

In the dense environment of GCs strong dynamical interactions are
expected, and these have often been proposed as the cause of sdB
formation through stellar collisions, merging, or encounters involving
binary stars, which as a consequence can harden \citep{Heggie75} and
eventually merge \citep{Bailyn92}.  In support of this hypothesis,
\citet{Fusi93} and \citet{Buonanno97} showed that more concentrated or
denser clusters tend to have bluer HB types and longer blue HB
``tails'', while \citet{Djorgovski92} observed a relation between the
dynamical history of GCs and their UV flux, a signature of a hot population
that could include sdBs. In some clusters
color gradients, with bluer centers, have been identified but,
mainly due to low number statistics because the light is
dominated by very few bright giants, it is not clear if
this feature is a general property of GCs with HB blue tails
\citep{Piotto88,Djorgovski01}

Nevertheless, other observational constraints put this
picture in doubt. In fact, one would expect to find EHB stars more
concentrated toward cluster centers where interactions are stronger,
but no radial gradient has been found in GCs with the most strongly
populated EHBs, such as NGC\,2808 \citep{Bedin00}, $\omega$~Cen
\citep{DCruz00}, NGC\,6388 and NGC\,6441 \citep{Rich97}.  On the other
hand, it should be noted that blue stragglers, which are also supposed
to be formed through stellar encounters, are indeed more concentrated
in the central regions of all GCs \citep{Piotto04}. Moreover, the
presence of pairs of clusters that are dynamically similar and yet
have very different HB morphology \citep[e.g.,
M\,3/M\,13][]{Ferraro97b}, or similar in HB morphology but dynamically
very different \citep[e.g., M\,15/NGC\,288][]{Crocker88} also argue
against dynamical interactions as an important channel for EHB star
formation.

\subsection{Results on EHB stars}
\label{introbservations}

As previously noted, there is a large difference between the apparent
magnitudes of field sdB stars and their GC counterparts. The nearest
sdBs in the Galactic field reach a visual magnitude $V=10-11$,
whereas in GCs they are much farther away, and are never brighter
than about $V=17$. To date, the study of EHB stars in GCs has been
limited to their photometric properties or low-resolution spectroscopy
in few GCs, and many results achieved on field sdBs have not been
feasible on EHB cluster stars yet. The most frequent targets of
investigations are the nearby GC NGC\,6752, and the very massive and
peculiar (due to their multimodal HB morphologies and multiple MS)
GCs $\omega$~Cen and NGC\,2808. As a matter of fact, $\omega$~Cen is not a
typical GC at all, and it has been repeatedly proposed as the nucleus
of a merged dwarf galaxy \citep*[][and references therein]{Altmann05},
resembling in this sense the case of M\,54
(NGC\,6715) in the Sagittarius dwarf
spheroidal galaxy \citep[][and references therein]{Layden00}.

One important feature in the CMD of many GCs is the often
underpopulated region separating sdBs from cooler HB
stars. \citet{Ferraro98} indicated the presence of a gap at about
$T_\mathrm{eff}=18,\!000$~K, while \citet{Piotto99} argued for a gap
at constant stellar mass, with $T_\mathrm{eff}$ varying with the
metallicity of the cluster \citep[see][for a more recent
discussion]{Momany04}.  The presence of this gap suggested to many
authors that sdB stars could be a distinct population with a different 
formation history compared to the other HB stars

\citet{Momany02} discovered that EHB stars hotter than
$T_\mathrm{eff}\approx 23,\!000$~K deviate from canonical tracks in
the CMD of NGC\,6752, appearing brighter than expected in
$U$. They argue that this feature could be due to the onset of
radiative levitation after the disappearance of the He\,{\sc ii}
convective layer, and this so-called ``Momany jump'' was later identified also
in other clusters \citep{Momany04}.  Unfortunately, no high-resolution
spectroscopic analysis has probed this phenomenon so far, and in
general there is still a lack of atmospheric abundance measurements
for EHB stars in GCs. Support for their hypothesis comes from
low-resolution observations by \citet{Moehler00}, who measured a very
low He abundance for some targets hotter than $23,\!000$~K
($\log{\mathrm{[He/H]}}\leq -3$), and a slight depletion for cooler EHB
stars. On the contrary, \citet{Moni07} failed to reproduce these
results, finding an almost constant He underabundance regardless of
temperature ($\log{\mathrm{[He/H]}}\approx -2$).  On the other hand,
among stars hotter than the Momany jump they find puzzling evidence of
two distinct groups of stars, with different photometric and
spectroscopic properties: the brighter stars are well reproduced by
canonical theoretical models, while the remainder show unphysically
high masses, which these authors attribute to a mismatch between models
and real stars.

Two classes of rapidly pulsating stars are known among field sdB stars
\citep{Kilkenny97,Green03}, and they are currently the subject of
extensive studies. Asteroseismological techniques are successfully used
to analyze their oscillations and derive their physical parameters
\citep{Randall05}. The detection of such pulsators in GCs is a technically
challenging issue, which however holds the promise to unveil the physical
mechanism(s) that lead to the production of the EHB (and blue hook) stars.
Surveys have proven fruitless until now \citep{Reed06}, but the search is not
hopeless and new projects have recently been undertaken \citep{Catelan07c}.

\subsection{Blue Hook Stars}
\label{introbluehook}

Over the past few years, 
the discovery of very faint EHB stars in the UV CMDs of $\omega$~Cen
\citep{Whitney98} and NGC\,2808 \citep{Brown01} has drawn 
considerable attentio.  These very hot stars
form a so-called ``blue hook'' at the blue end of the HB, and are
hotter than the canonical theoretical limit for EHB stars. They were
proposed as the progeny of stars which -- due to extreme mass
loss -- did not ignite He at the tip of the RGB, but only
later along the WD cooling sequence \citep[``late hot flashers'',
e.g.,][]{Castellani93,DCruz96,Brown01}. Spectroscopically measured
temperatures and He abundances are consistent with the outlined
picture \citep{Moehler02,Moehler04}. Detailed calculations showed
that EHB stars formed through a late He flash should show higher He
and C abundances \citep{Cassisi03,Lanz04} as a consequence of mixing
events during the flash, at variance with the primordial He
enhancement scenario for which no C overabundance is expected and He
should not exceed the primordial (although enhanced) value.  Detailed
spectroscopic observations should therefore be able to tell between
the competing hypotheses. Very recent results by \citet{Moehler07}
strongly support the late-flash scenario. Blue hook stars have also been
found in NGC\,6388 and NGC\,6441 \citep{Busso07}, NGC\,2419
\citep{Ripepi07}, and M\,54 \citep{Rosenberg04,Momany04,Siegel07}, and a star
spectroscopically studied in M\,15 \citep{Moehler95} is now recognized
as a possible blue hook candidate on the basis of its temperature and
He abundance; however, blue hooks are completely lacking in many other
clusters.  In NGC\,6752, for example, the EHB ends at temperatures
adequately predicted by standard models, and no blue hook extension is
observed. If these stars are the progeny of late flashers, it is
unclear why they occur only in the most massive GCs \citep[the
clusters named above are among the most massive in the Galaxy; see
also] [for recent evidence pointing to cluster mass as a strong
second-parameter candidate]{Recio06}.

\subsection{Binary sdB Stars}
\label{introbinaries}

Binary interactions are supposed to play a major role in the formation
of sdB stars, as first proposed by \citet{Mengel76}.
The most recent models of binary evolution provide a satisfactory
explanation for both the origin and observed properties of these stars
\citep{Han02,Han03}. Models in which sdBs arise from binary
interaction have also been used to successfully model the observed
UV upturn in external galaxies \citep{Han07}, which is characterized
by an increase in UV flux beyond the predictions of canonical population
synthesis models, and which is likely to originate from a sdB star 
population \citep[see for example][ and Catelan 2007a for a recent
review]{Greggio90}.

On the observational side, numerous surveys have pointed to the
presence of a large population of binaries among field sdB stars,
although the results often disagree on the exact binary
fraction, probably because of selection effects that affect the
samples to different extents. Early investigations focused on the IR
flux excess arising from a MS companion unseen at optical wavelengths,
and estimated that 50-66\% of field sdBs reside in sdB+MS systems
\citep{Ferguson84,Allard94,Ulla98,Aznar01}. These surveys brought no
information about binary periods, were blind to compact
companions, and limited to a lower limit in MS companion
mass for the detection. More recently \citet{Reed04} found a much
lower fraction for sdB+MS systems (20\%), and they argued that
previous results were strongly affected by biased samples.  Later
studies relied on radial velocity (RV) measurements for binary
detection.  They are more sensitive to short-period systems and
higher-mass companions, but searches for wide binaries are ongoing
\citep{MoralesRueda06}.  \citet{Maxted01} found many RV variables
among their sdB targets, mainly with unseen white dwarf companions
(sdB+WD binaries), and estimated that nearly 70\% of sdBs are close
systems.  \citet{Napiwotzki04} found a much lower but still fairly
high fraction (about 40\%), and many investigations of orbital
parameters pointed out that close binaries with period $P\leq 10$~d
are very common among field sdB stars
\citep{Moran99,Saffer98,Heber02,MoralesRueda04}: in fact, such systems
are suspected to constitute nearly half of the total Galactic sdB
population.

The success of theoretical models and the results of observational
investigations have led to the generally accepted conclusion that sdB
stars are strictly related to binary interactions, with close
systems playing a major role amongst them. It is only over the past
few years that similar investigations have started being carried out
in globular star clusters as well. These have led to some dramatic, 
and indeed quite unexpected, results.


\section{sdB Binaries in GCs: previous Results}
The first results of a search for close binaries among EHB stars in
GCs were presented in the second meeting of this series
\citep{Moni06c}. The analysis of NGC\,6752 observations revealed a striking
lack of close binary systems: indeed, no close binary was then detected
in a sample of 18 targets with $T_\mathrm{eff}\geq20,\!000$~K. As surveys
usually span a large range of temperatures along the HB, this temperature
limit will be implicitly assumed to define EHB stars hereafter.
The survey was optimized to detect close systems with period
$P\leq 10$~d, and it was almost blind to longer periods and/or sdBs with
low-mass companions \citep[typically about $0.1\, M_{\odot}$, as in the case
of the well-known HW Vir system;][]{Menzies86}. From these results
\citet{Moni06a} calculated that, at the 95\% confidence level, the close
binary fraction among EHB stars in NGC\,6752 is lower than 20\%.
In the same cluster \citet{Peterson02} had claimed
the detection of many radial velocity (RV) variables; however, the reasons for
this discrepancy remain unclear \citep[see][for a discussion]{Moni06a}.

Later \citet{Moni06b} presented preliminary results of a similar work on
two new clusters, namely M\,80 (NGC\,6093) and NGC\,5986. Unfortunately they
are far from being as conclusive as the previous ones. In the case of M\,80,
they detected one good binary candidate out of 11 targets, and derived
a most probable close binary fraction $f_{P \leq 10\,{\rm d}} =11\%$,
but with large uncertainties because bad weather limited the extent of
the observations and accordingly lowered the sensitivity of the survey.
On the other hand, in the case of NGC\,5986 the temporal coverage was
good, leading to the detection of only one (doubtful) candidate; however,
the limited sample size (5 targets) prevented a full statistical analysis
of the results from being carried out.

It is worth noting that \citet{Moni07} individuated in NGC\,6752
an EHB star that behaves as an sdB+MS binary (star \#5865),
being redder than other targets at the same temperature and showing a strong
Mg~{\sc i}b triplet, which is most unusual for these stars.

The picture resulting from these investigations points towards a
noticeable lack of close binaries among EHB stars in GCs
compared with field sdBs. Many questions have arisen from these results
that still await answers based on suitable observational material. 
In particular, a more precise
estimate of $f$ in NGC\,6752 is needed in order to better determine how
significant the lack of binaries in this cluster is; in addition, it
must still be clarified whether a low $f$ 
is a common feature among GCs.
Finally, the role of wide binaries and/or low-mass companions has yet to
be investigated in detail.


\begin{figure}[!t]
\plottwo{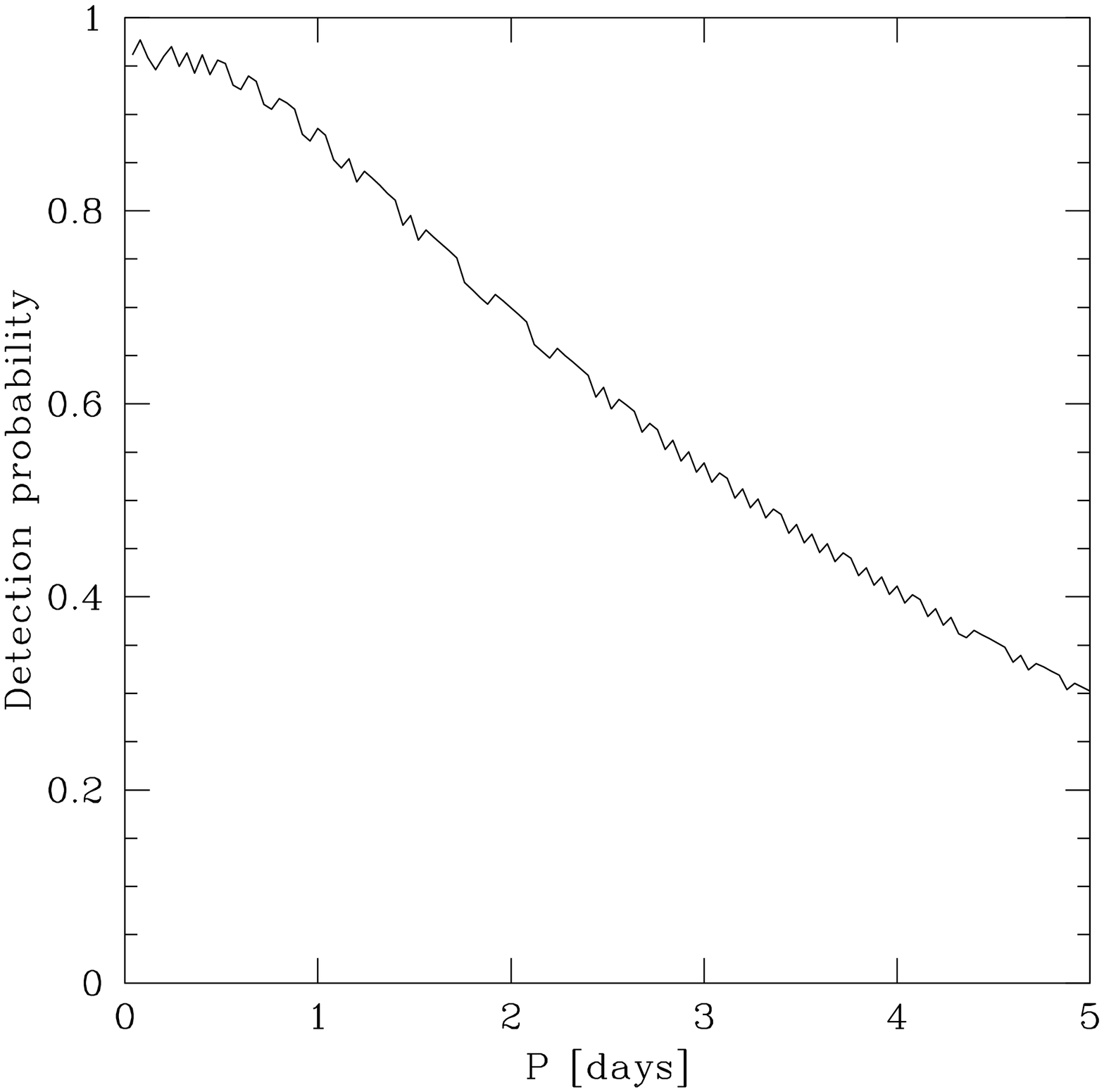}{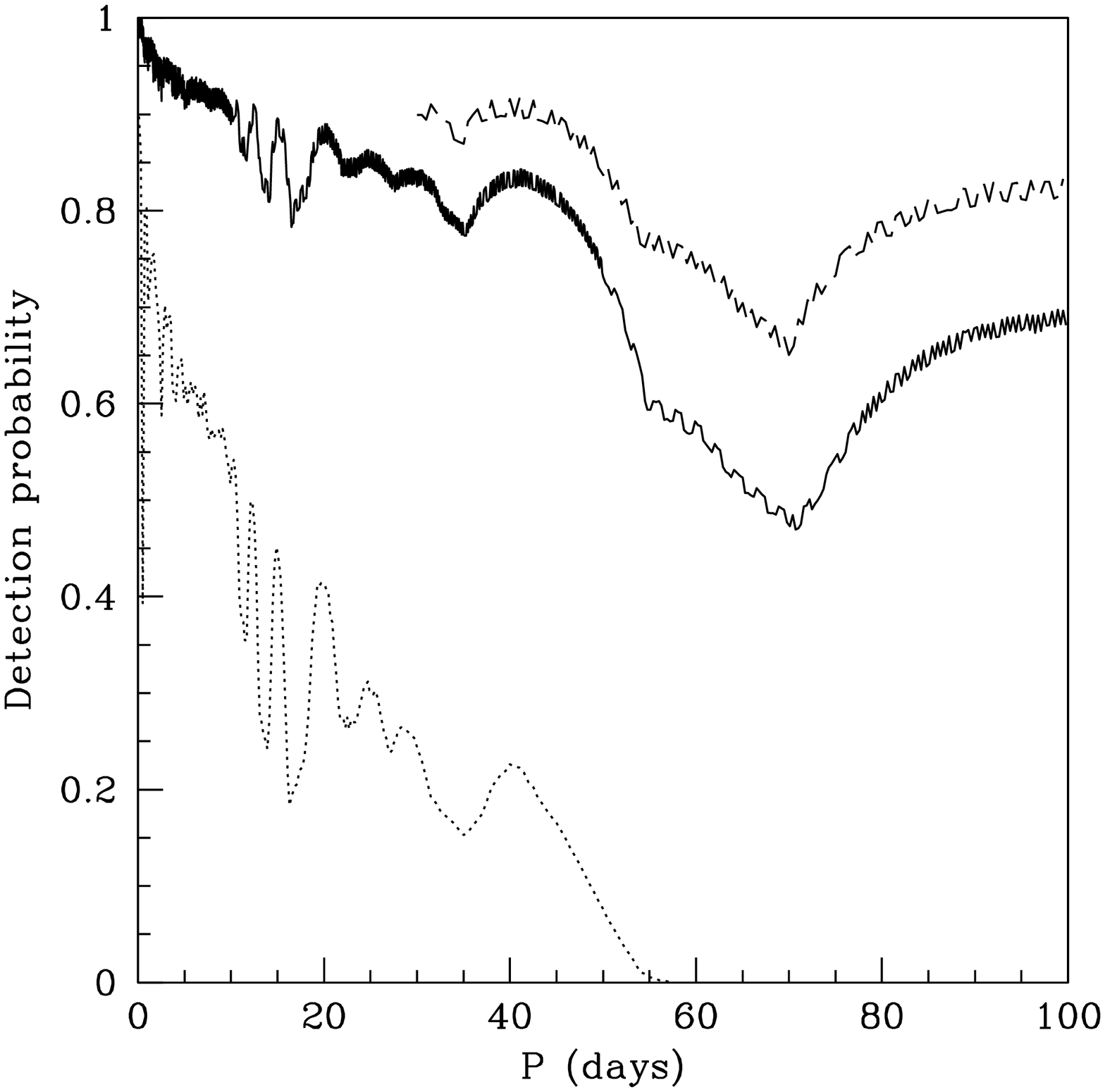}
\caption{Binary detection probability as a function of period, calculated
as in \citet{Moni06a}.  {\it Left panel:}
Probability for the NGC\,6752 survey, assuming a $0.5\,M_{\odot}$
mass companion.  {\it Right panel:} Probability for the
NGC\,2808 survey. The solid and dotted lines are calculated under 
the assumptions of a $0.5\,M_{\odot}$ and a $0.1\,M_{\odot}$
companion, respectively, and a $3\sigma$ threshold of $10\,{\rm km}\,{\rm s}^{-1}$. 
The dashed line refers to the case of a wide, $0.5\,M_{\odot}$ companion,
and a $3\sigma$ threshold of $6\,{\rm km}\,{\rm s}^{-1}$. }
\label{Figprob}
\end{figure}

\section{New Results on NGC\,6752}
Recently, in one observing night at VLT-UT2, we collected four
high-resolution spectra ($R=18,\!000$) for 54 EHB stars in NGC\,6752,
using the FLAMES-GIRAFFE spectrograph (setup H7A). Because of the fact
that the observations were limited to only one night, the survey is
sensitive only to the shortest periods (see Fig.~\ref{Figprob}). For
this reason, we restrict our analysis to $P\leq5$~d, for which we have
a detection probability higher than 40\%. We consider this to be a
minor problem, since the bulk of known close-binary sdB systems peaks
around $P=1$~d in the field, systems with $P\geq5$~d merely
constituting the tail end of the period distribution \citep[see, for
example,][]{MoralesRueda03}.

Radial velocities were measured by cross-correlating \citep{Tonry79} the
H$_\beta$ line with full wings.  The adopted synthetic template was
chosen from the library of \citet{Munari05}. We explicitly avoided
weaker metallic lines, because their quite unpredictable shape in our
high-resolution, low-S/N spectra resulted in distorting the
cross-correlation function without adding any information about the
position of its peak. The errors in the RV variations were calculated as
the quadratic sum of errors in the single RV measurements, but by accurate
analysis of cooler stars included in the sample we found that this 
procedure underestimated the error by a factor of 1.44. We applied such
a correction factor to obtain the final errors. Thirteen targets were 
excluded for different reasons (light contamination from nearby lamp 
fibers, one or more spectra with too low S/N, star badly centered in 
the fiber).

\begin{figure}
\plotone{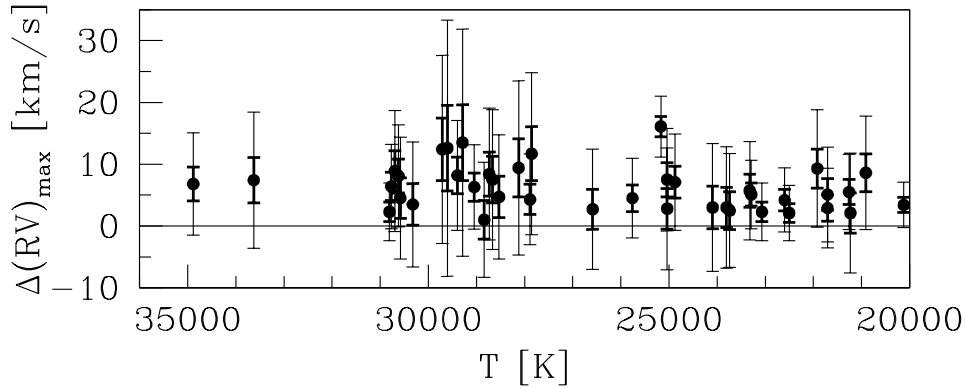}
\caption{Results for NGC\,6752. The absolute value of the maximum RV
variation for each star is plotted against the effective temperature,
evaluated on the basis of the color-temperature relation from \citet{Moni06a}.
The thick error bar is the $1\sigma$ value, while the thinner one is
the $3\sigma$ threshold.}
\label{6752res}
\end{figure}

Our results are shown in Fig.~\ref{6752res}, where we plot the maximum
RV variation as a function of the effective temperature.
The $3\sigma$ threshold is on average $10\,{\rm km}\,{\rm s}^{-1}$.
We detect only one target (at $T_\mathrm{eff}\approx 25,\!000$~K) showing a
variation greater than this threshold ($16.1\pm 1.6\,{\rm km}\,{\rm s}^{-1}$).
This star is the same one (\#5865) for which \citet{Moni07} find evidence
for a cool MS companion. Hence, it clearly is a sdB+MS close system, the
first one ever discovered in a GC.

With the detection probability plotted in Fig.~\ref{Figprob} and
assuming the detection
of one binary system out of 41 targets, we performed a statistical
calculation as in \citet{Moni06a}. We assumed both a flat period
distribution and a Gaussian one in $\log{P}$, but the resulting
differences are negligible.  The best estimate (most probable value) of
the close binary fraction is $f_{P\leq 5\,{\rm d}} =4\%$. We also find
that, at the 95\% confidence level, $f_{P\leq 5\,{\rm d}} \leq 16\%$,
in good agreement with previous results for this cluster.


\section{NGC\,2808: Preliminary Results}
We are performing an extensive search for binaries among EHB stars in
the globular cluster NGC\,2808, using the same instrument and methods
as in the investigation presented in the previous section.

In NGC\,2808 we observed 27 targets twice in each of the six observing
half-nights distributed in the course of 71 days. The survey is optimized 
to search for binaries of any expected type.
The $3\sigma$ threshold is about $10\,{\rm km}\,{\rm s}^{-1}$, and given
the good temporal distribution of observations the resulting detection
probability is very high ($\geq 80\%$) for periods $P\leq 40$~days
(see Fig.~\ref{Figprob}). For
wider binaries the sensitivity drops slowly, but they are not
expected to show noticeable variations within a 4\,h interval, hence we
can sum the two spectra collected each night and reach a
$3\sigma = 6\, {\rm km}\,{\rm s}^{-1}$ threshold. The probability of detecting
binaries with low-mass companions is also high for the
shortest periods ($P \leq 10$~days) observed among field sdBs.

Here we present some preliminary results of this ongoing project.  The
analysis is still in a first stage, so the present discussion merely provide a hint
into what the final results may turn out to be. We analyzed the data
collected in the first 4 nights for 6 EHB targets, spanning a temporal
interval of 6 days. Hence we are still dealing with close binaries
only. We find two stars showing variations greater than $3\sigma$. One
of them (around $21,\!000$~K) needs further accurate study because
variations are neither high nor far from the $3\sigma$ threshold. The
hotter one at about $24,\!000$~K, on the other hand, shows large
variations that can hardly be explained as the result of inaccuracies
in our analysis. A seventh star currently under study seems to
present a similar behavior.

This preliminary overview of our data suggests that, at variance
with other clusters studied so far, close binaries could be
relatively common among the NGC\,2808 EHB star population.
It is worth considering that the sampled analyzed so far is small,
and the statistic is still too poor.


\section{Discussion}

With our new investigation we confirm that close binaries are indeed
very rare among EHB stars in NGC\,6752.  Our estimated upper limit for
the close binary fraction agrees well with previous results, and
the most probable value is small (4\%).  The difference between
field sdB stars and their analogs in this cluster is evident.
Still nothing can be said about other types of binaries found
among field sdBs, i.e. wide binaries and short-period sdBs with very
low-mass companions. We are working to fill this gap, but 
investigations of more clusters are also needed.

There are indications that close EHB systems are rare also in
two other GCs investigated so far, namely M\,80 and
NGC\,5986. This points to a difference between the formation of sdB 
stars inside and outside GCs, and any successful model
must explain this observational constraint.

Close binary sdB stars are considered the progeny of systems that
underwent one or two common envelope (CE) phases \citep{Han02}. 
Accordingly, this mechanism should not be efficient inside GCs,
contrary to what is observed in the field. Our present 
knowledge still cannot clarify the role of other binary channels that 
are theoretically expected to produce sdBs, such as the stable Roche
Lobe Overflow (RLOF) channel (which produces wide-orbit binaries) and 
the merger channel (whose progeny are single sdB stars).

Moni Bidin et al. \citetext{2007, submitted} propose that among
different sdB populations an $f$-age relation should exist, because
the CE channel should be less efficient with increasing age. Detailed
searches for long-period EHB binaries could test their hypothesis, but
wide binaries could also be lacking, if either the merger channel or 
single-star evolution are the main sdB formation mechanisms inside GCs.
In fact, the merger channel should be more efficient for old populations
\citep[][their Fig.~7]{Han07}, and it could also be enhanced by stellar
interactions \citep{Bailyn89}. Wide binaries, on the other hand,
can be easily disrupted in dense environments \citep{Heggie75}.
In any case, \citet{Moni06a} point out that the lack of a radial gradient
for EHB stars in GCs (as discussed in \S1) puts their binary
origin in doubt, because typical systems as observed in the field should
exceed the turnoff mass (about $0.7-0.8\, M_\odot$), and thus migrate
toward the central regions due to mass segregation effects. Moreover,
we would expect that members of disrupted systems or sdBs formed
through stellar encounters and collisions would be a kinematically
hotter population, but in our data we find no evidence of a difference
with respect to other HB stars.

It is well known that even MS and RGB binaries are much more
frequent in the field than in GCs
\citep[see for example][for a recent study]{Sollima07}, but this
would not suffice to account for the difference in EHB binary fraction,
as it would imply that sdB formation and binarity are unrelated~-- at variance
with our present understanding of these stars.
Moreover, the binary fraction among MS field stars could even be noticeably
lower than has been assumed so far \citep{Lada06}.

NGC\,2808 could be a remarkably different case. If our preliminary results 
are confirmed, this cluster would be relatively rich in close 
EHB binaries, in contrast to the other clusters studied so
far. It is tantalizing to relate this discrepancy to
other well-known peculiarities of the cluster, as its multimodal HB
\citep{Sosin97} or its triple MS \citep{Piotto07}, but
our study is still in too early a stage and the sample analyzed is too small
to allow definitive conclusions.


\acknowledgements   CMB's attendance to the meeting was funded by program
Fundaci\'on Andes C-13798 and the LOC. MA's attendance was funded by the LOC.
CMB warmly thanks the LOC for constant assistance during the meeting.
MC is supported by Proyecto Fondecyt Regular \#1071002.

\bibliographystyle{apj}

\bibliography{Moni_Bamberg4}

\end{document}